\documentclass[twocolumn,showpacs,aps,prl,superscriptaddress,floatfix]{revtex4}

\usepackage{graphicx}
\usepackage{dcolumn}
\usepackage{amsmath}
\usepackage{epsfig}

\input pubboard/babarsym
\newcommand{\deltae}{\ensuremath{\Delta E}}

\def\Dp      {\ensuremath{D^+}}
\def\Dm      {\ensuremath{D^-}}
\def\Dstar   {\ensuremath{D^{*}}}
\def\Dstarp  {\ensuremath{D^{*+}}}
\def\Dstarm  {\ensuremath{D^{*-}}}

\def\rhop    {\ensuremath{\rho^+}}

\def\aonep   {\ensuremath{a_1^+}}

\newcommand{\BABARPubYear}    {01}
\newcommand{\BABARPubNumber}  {02}

\newcommand{\SLACPubNumber} {9061}

\def\figurebox#1#2#3{%
    \def\arg{#3}%
    \ifx\arg\empty
    {\hfill\vbox{\hsize#2\hrule\hbox to #2{\vrule\hfill\vbox to #1{\hsize#2\vfill}\vrule}\hrule}\hfill}%
    \else
    {\hfill\epsfbox{#3}\hfill}%
    \fi}

\providecommand{\hips}{\mbox{\ensuremath{\ps^{-1}}}}
\long\def\inst#1{\par\nobreak\kern 4pt\nobreak
    {\it #1}\par\vskip 10pt plus 3pt minus 3pt}

\begin{document}

%\preprint{HEP/123-qed}
\begin{flushleft}
\babar-PUB-\BABARPubYear/\BABARPubNumber\\
SLAC-PUB-\SLACPubNumber\\[10mm]
%hep-ex/\LANLNumber\\[10mm]
\end{flushleft}

\title[Short Title]{\large \bf
 Measurement of {\boldmath{\Bz-\Bzb}} Flavor Oscillations in
Hadronic {\boldmath \Bz} Decays }

%% author list as of 16-Nov-2001 (549 authors)
%
\author{B.~Aubert}
\author{D.~Boutigny}
\author{J.-M.~Gaillard}
\author{A.~Hicheur}
\author{Y.~Karyotakis}
\author{J.~P.~Lees}
\author{P.~Robbe}
\author{V.~Tisserand}
\affiliation{Laboratoire de Physique des Particules, F-74941 Annecy-le-Vieux, France }
\author{A.~Palano}
\author{A.~Pompili}
\affiliation{Universit\`a di Bari, Dipartimento di Fisica and INFN, I-70126 Bari, Italy }
% ...
\collaboration{The \babar\ Collaboration}
\noaffiliation

\date{December 31, 2001}

\begin{abstract}
Flavor oscillations of neutral \B\ mesons have been studied in
\epem\ annihilation data collected with the
\babar\ detector at center-of-mass energies near the \FourS\ resonance.
The data sample used for this purpose consists of events in which 
one \Bz\ meson is reconstructed 
in a hadronic decay mode,
while the flavor of the recoiling \Bz\ is determined with a tagging
algorithm that exploits the correlation between the flavor of the
heavy quark and the charges of its decay products.
From the time development of the
observed mixed and unmixed final states we determine
the \Bz-\Bzb oscillation frequency \deltamd\ to be 
$0.516 \pm 0.016 ({stat}) \pm 0.010 ({syst})\,\hips$.

\end{abstract}

% PACS, the Physics and Astronomy Classification Scheme.

\maketitle

In the Standard Model,
\Bz-\Bzb mixing~\cite{conjugate} occurs through second-order weak diagrams
involving the exchange of up-type quarks,
with the top quark contributing
the dominant amplitude.  A measurement of the mass difference
\deltamd between the mass eigenstates is therefore
sensitive to the value of the Cabibbo-Kobayashi-Maskawa~\cite{KM} 
matrix element $V_{td}$.
The phenomenon of particle--anti-particle mixing
in the neutral \B\ meson system was first seen almost fifteen years
ago~\cite{B0mix}.
The oscillation frequency \deltamd\ has been 
%extensively studied 
measured with both time-integrated  and
time-dependent techniques~\cite{PDG2000}.

In this Letter we present a measurement of time-dependent mixing
based on a sample of 32 million \BB\ pairs recorded
at the \FourS\ resonance with the \babar\ detector at the Stanford Linear
Accelerator Center. 
This study and a related \CP\ asymmetry measurement~\cite{BabarPub0118}
are described in more detail
in Ref.~\cite{BabarPub0103}.
At the PEP-II asymmetric-energy $e^+e^-$ collider,
the \FourS\  provides a
source of $\Bz\Bzb$ pairs moving along the $e^-$ beam direction ($z$ axis)
with a known Lorentz boost of $\beta\gamma = 0.55$, allowing a novel technique
for determining \deltamd.

The \Bz-\Bzb\ mixing probability, for a given \Bz\ lifetime $\tau$,
is a function of \deltamd\ and the proper
decay-time difference \deltat\ between the two neutral $B$ mesons produced
in a coherent $P$-wave state in the
\FourS event.
The result is a time-dependent probability to observe {\em unmixed} $(+)$,
$\BzBzb$, 
or {\em mixed} $(-)$, $\Bz\Bz$ and $\Bzb\Bzb$, events:
\begin{eqnarray}
Prob(\Bz\Bzb \rightarrow \Bz\Bzb, \Bz\Bz\ \mbox{\textrm or\ }\Bzb\Bzb) \propto & & \nonumber \\
{\rm{e}}^{-|\deltat|/\tau} (1 \pm \cos \deltamd \deltat). & &
\label{eq:mix} 
\end{eqnarray}
The effect can be measured by reconstructing
one \B\ in a flavor eigenstate, referred to as $B_{\rm rec}$, 
while the remaining
charged particles originating from
the decay of the other \B, referred to as $B_{\rm tag}$, 
are used to identify, or ``tag'', its flavor as a \Bz\ or \Bzb.
The charges of identified leptons and kaons are the primary indicators, 
although other information in the event can also be used to 
identify the flavor of $B_{\rm tag}$. The time difference 
$\deltat = t_{\rm rec} - t_{\rm tag}\simeq \deltaz/\beta\gamma c$ 
is determined from the separation $\deltaz$ of the decay vertices
for the flavor-eigenstate and tagging \B\ decays
along the boost direction.
The tagging and vertexing algorithms used in this analysis 
are nearly identical to those
employed for \CP\ violation studies,
in which one \B\ is fully reconstructed in a
\CP\ eigenstate~\cite{BabarPub0118}. 

The value of \deltamd\ is extracted from a tagged 
flavor-eigenstate \Bz\ sample with a simultaneous unbinned
maximum likelihood fit to the \deltat\ distributions of
mixed and unmixed events. There are two principal experimental
factors that complicate the
probability distribution 
given by Eq.~(\ref{eq:mix}).
First, the tagging algorithm, which classifies events into
categories $i$ depending on the source of the available tagging
information, incorrectly identifies the 
flavor of $B_{\rm tag}$ with a probability $\mistag_i$. This mistag 
rate reduces the observed amplitude of the oscillation 
by a factor $(1 -2\mistag_i)$.
Second, the resolution for 
\deltat\ is  comparable to the oscillation period
and must be well understood.
The probability density functions (PDFs)
for the unmixed and mixed signal events, 
${\cal H}_{\pm, {\rm sig}}$, can be expressed as
the convolution of the underlying 
\deltat\ distribution for the $i^{th}$ tagging category,
\begin{equation*}
h_\pm(\deltat; \deltamd, \mistag_i)  = \frac{{\rm e}^{ -\left| \deltat \right|/\tau}}{4\tau}
\left[ 1 \pm (1-2\mistag_i)\cos{ \deltamd \deltat } \right],
\end{equation*}
with a \deltat\ resolution
function 
containing parameters $\hat {a}_j$.
A log-likelihood function is then constructed 
by summing
$\ln{\cal {H}}_{\pm, {\rm sig}}$ over all 
events within each of the 
tagging categories.
The likelihood is maximized to extract simultaneously the mistag 
rates $\mistag_i$,
the resolution function parameters $\hat{a}_j$, and the 
mixing parameter \deltamd.

The \babar\ detector is
described in detail elsewhere~\cite{BabarPub0108}.
Charged particles are detected and their momenta  measured by a combination of
a 40-layer drift chamber (DCH) and a five-layer silicon vertex tracker (SVT) embedded
in a 1.5-T solenoidal magnetic field.
$B_{\rm rec}$ decay
vertices are reconstructed with a resolution of typically
65\mum\ along the boost direction. 
A detector of internally reflected
Cherenkov radiation (DIRC)
is used for charged hadron identification.
Kaons are identified with a neural network based on the
likelihood ratios in the SVT
and DCH, derived from \dedx\ measurements, and in the DIRC,
calculated by comparing the observed and
expected pattern of Cherenkov light for either kaons or pions.
A finely segmented CsI(Tl)
electromagnetic calorimeter (EMC) is used to
detect photons and neutral hadrons, and to identify electrons.
Electron candidates are required to have a ratio
of EMC energy to track momentum, an EMC cluster shape, 
DCH \dedx, and DIRC Cherenkov angle consistent with expectation.
The instrumented flux return (IFR) is segmented and
contains
resistive plate chambers for muon
and neutral hadron 
identification. Muon candidates are required to have IFR hits
located along the extrapolated DCH track, an IFR penetration length,
and an energy deposit in the EMC consistent with the muon hypothesis.

Neutral \B\ mesons are reconstructed in a sample of multihadron
events in the flavor eigenstate decay modes $D^{(*)-}\pip$,
$D^{(*)-}\rho^+$, $D^{(*)-}a_1^+$ and $\jpsi\Kstarz$.
The decay channels $\Kp\pim$, $\Kp\pim\piz$, $\Kp\pip\pim\pim$ and
$\KS\pip\pim$ are used to reconstruct $\Dzb$ candidates, while
the modes $\Kp\pim\pim$ and $\KS\pim$ are used for $\Dm$ candidates. 
Charged $\Dstarm$ candidates are formed by combining
a $\Dzb$ with a soft $\pim$. Finally, the \Bz\ candidates are
formed by combining 
a \Dstarm\ or \Dm\ candidate
with a \pip, \rhop\ $(\rhop\to\pip\piz)$ or \aonep\ $(\aonep\to\pip\pim\pip)$;
likewise,
$\Bz\to\jpsi\Kstarz$ candidates are reconstructed
from combinations of \jpsi\ candidates, in the decay modes
$\epem$ and $\mumu$, with a \Kstarz\ $(\Kstarz\to\Kp\pim)$.
The selection and reconstruction of these decay chains and the
selection of multihadron events, including continuum supression,
is described in more detail in Ref.~\cite{BabarPub0106}.

Neutral \B\ candidates are identified by the
difference \deltae\ between the energy of the candidate and the
beam energy $\sqrt{s}/2$ in the center-of-mass frame,
and the beam-energy substituted mass \mes, calculated from $\sqrt{s}/2$
and the reconstructed momentum of the $B$ candidate.
Candidates are selected by requiring
$\mes>5.2$\gevcc\ and $\Delta E$ within $\pm 2.5$ standard deviations of 0
(typically $\left| \Delta E \right| < 40$\mev).
When multiple candidates in a given event are selected (with probability
of about 0.25\%),
only the one with the smallest $|\deltae|$ is retained.

After the daughter tracks of the $B_{\rm rec}$ are removed,
the remaining tracks are analyzed to determine the flavor of
the $B_{\rm tag}$.
For this purpose, we use the flavor tagging information carried by 
primary leptons from semileptonic $B$ decays, charged kaons, soft pions from
\Dstar\ decays, and high momentum charged particles, to
assign an event to a single tagging category.

Events are assigned a {\tt Lepton} tag
if they contain an identified lepton
with a center-of-mass momentum greater than
1.0 or 1.1\gevc\ for electrons and muons, respectively.
The momentum requirement selects mostly primary leptons by suppressing
opposite-sign leptons from semileptonic charm decays.
If the sum of charges of all identified kaons is non-zero,
the event is assigned a {\tt Kaon} tag.
The final two tags involve a multivariable analysis based on a
neural network, which is trained
to identify primary leptons, kaons, and soft pions, and
the momentum and charge of the track with the maximum center-of-mass
momentum.  Depending on the output of the neural net, events are
assigned either an
{\tt NT1} (more certain)  tag, an {\tt NT2} (less certain) tag, or are not
tagged at all (about 30\% of all events) and excluded from the analysis.

Tagging assignments are made mutually exclusive by the hierarchical
use of the tags. Events with a {\tt Lepton} tag and no conflicting {\tt Kaon}
tag are assigned to the {\tt Lepton} category. If no {\tt Lepton} tag exists,
but the event has a {\tt Kaon} tag, it is assigned to the {\tt Kaon} category.
Otherwise events with neural network tags are assigned 
to corresponding neural network categories.

The decay time difference \deltat\ between \B\ decays is determined from the
measured separation $\Delta z = z_{\rm rec}-z_{\rm tag}$
along the $z$ axis between the reconstructed $B_{\rm rec} (z_{\rm rec})$ and
flavor-tagging decay $B_{\rm tag}$ $(z_{\rm tag})$ vertex.
This measured $\Delta z$ is converted into \deltat with the use of the
known \FourS\ boost, including a
correction on an event-by-event basis for the direction of the \B\ mesons
with respect to the $z$ direction in the \FourS\ frame.
The \deltat\ resolution is dominated by the $z$ resolution 
of the tag vertex position. 
Reconstruction of the $B_{\rm tag}$ decay vertex starts with all
tracks in the event except those incorporated in $B_{\rm rec}$. In order to
reduce the contamination from $D$ meson decay products, those identified 
as kaons are also excluded.
An additional constraint is provided by the calculated
$B_{\rm tag}$ production point and three-momentum, determined
from the 
momentum of the $B_{\rm rec}$ candidate, its decay vertex,
the average position of the interaction point, and the \FourS\ boost.
Tracks with a large contribution to the $\chi^2$
are iteratively removed from the fit
until those remaining $(\geq 1)$ have a reasonable fit probability
or all tracks are removed.
Only events with a reconstructed $B_{\rm tag}$ vertex,
$|\deltat|<20$\ps\ and $\sigma_{\deltat} < 1.4$\ps are
retained (about 84\%),
where $\sigma_{\deltat}$ is the measurement error derived from the
vertex fits.

\begin{figure}[!tbh]
\begin{center}
 \includegraphics[width=\linewidth]{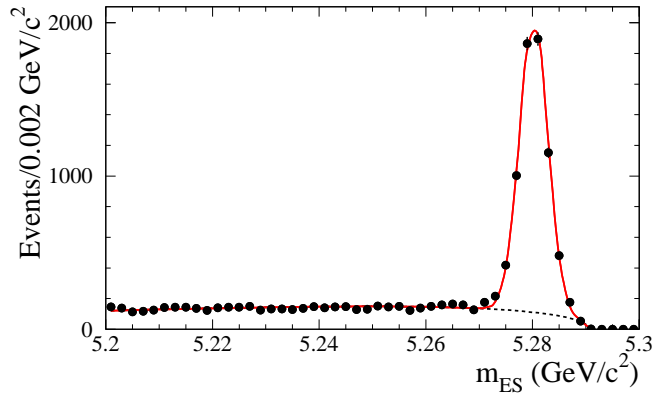}
\end{center}
\caption{Distribution of \mes\ for all \Bz\ candidates 
with a flavor tag and a reconstructed tag vertex.
\label{fig:hadronicb0}}
\end{figure}

The distribution of \mes\ for the selected candidates is shown
in Fig.~\ref{fig:hadronicb0}, where the result of a fit with
a Gaussian distribution for the signal and an ARGUS
function~\cite{ARGUS_bkgd} for the background is also displayed.
The fitted number of signal events and their purity (for $\mes > 5.27\gevcc$) 
are $6347\pm 89$ and $(85.8\pm 0.5)$\%, respectively.
The sample composition
by tagging category is given in Table~\ref{tab:HadronicBYield}.

\begin{table}[!htb]
\caption{
Signal yields per tagging category,
obtained from the \mes\ distributions after all selection
requirements.  The purity is quoted for $\mes >5.27 \gevcc$.}
\begin{center}
\begin{tabular}{lcc} \hline\hline
Category     & Tagged        & Purity (\%) \\ \hline
{\tt Lepton} & $1097\pm 34$   & $96.0\pm 0.7$  \\
{\tt Kaon}   & $3156\pm 63$   & $84.6\pm 0.7$  \\
{\tt NT1}    & $\;\, 798\pm 31$   & $88.9\pm 1.2$  \\
{\tt NT2}    & $1293\pm 43$   & $79.4\pm 1.3$  \\ \hline
\hline
\end{tabular}
\end{center}
\label{tab:HadronicBYield}
\end{table}

In the likelihood fit, the \deltat\ resolution function
is approximated by a sum of three Gaussian distributions (core, tail, and outlier) with different means and widths,
\begin{eqnarray}
{\cal {R}}( \delta_{\rm t} ; \hat {a} ) &=&  \sum_{k=1}^{2} 
{ \frac{f_k}{S_k\sigma_{\deltat}\sqrt{2\pi}} \, {\rm exp} 
\left(  - \frac{( \delta_{\rm t}-b_k\sigma_{\deltat})^2}{ 
 2({S_k\sigma_{\deltat}})^2 }  \right) } + \nonumber \\
&& { \frac{f_3}{\sigma_3\sqrt{2\pi}} \, {\rm exp} 
\left(  - \frac{ \delta_{\rm t}^2}{ 
 2{\sigma_3}^2 }  \right) },\nonumber
\label{eq:vtxresolfunct}
\end{eqnarray}
where $\delta_{\rm t}=\deltat-\deltat_{\rm true}$. The sum of the fractions
$f_k$ is constrained to unity.
For the core and tail Gaussians, the widths
$\sigma_{k}=S_{k}\times\sigma_{\deltat}$ are the event-by-event
measurement errors scaled by a common factor $S_{k}$.
The scale factor of the tail Gaussian is fixed to the Monte Carlo value
since it is strongly correlated with the other resolution function
parameters.
The third Gaussian, with a fixed width of $\sigma_3=8$\ps, accounts for
outlier events with incorrectly reconstructed vertices (less than 1\% of events).
A separate core bias coefficient $b_{1,i}$ is allowed
for each tagging category $i$ to account for small
shifts due to inclusion of charm decay products in the tag vertex,
while a common bias coefficient $b_2$ is used for the
tail component. These offsets are proportional
to $\sigma_{\deltat}$ since both the size of the bias 
and the resolution for $z_{\rm tag}$
depend kinematically on the polar angle of
the flight direction of the charm daughter.
The tail and outlier fractions and the scale factors
are assumed to be the same for all decay modes,
since the $z_{\rm tag}$ measurement dominates the resolution for \deltat.
This assumption is confirmed by Monte Carlo studies.
Separate resolution parameters are used for two different data-reconstruction
periods, referred to as {\tt Run1} and {\tt Run2}, 
which mainly differ in vertex performance
and tracking efficiency.

In the presence of backgrounds, additional terms must be added to the 
signal PDF ${\cal H}_{\pm, {\rm sig}}$ for each background source,
\begin{equation}
{{\cal H}_{\pm,i}} =
 f_{i, {\rm sig}}{\cal H}_{\pm, {\rm sig}} +
 \sum_{j={\rm bkgd}} f_{i,j} {\cal{B}}_{\pm,i,j}(\deltat;\hat b_{\pm,i,j}), \nonumber
\end{equation}
where
the background PDFs ${\cal{B}}_{\pm,i,j}$ provide an empirical
description for the \deltat\ distribution of the background events 
in each tagging category $i$ of the sample.
The fraction of background events for each tagging category 
and background source is given by
$f_{i,j}$, while $\hat b_{\pm,i,j}$ are parameters used to characterize each
source of background by tagging category for mixed and unmixed events.  
The signal probability $f_{i, {\rm sig}}$ 
is determined from the measured event \mes on the basis of a separate 
fit to the observed \mes\ distribution in
tagging category $i$.
The sum of signal and background fractions is forced to unity.

The \deltat\ distributions of the 
background are described with a zero lifetime component and a
non-oscillatory component with non-zero lifetime.
We fit for separate resolution function
parameters for the signal and the background 
to minimize correlations.
The likelihood
fit involves a total of 44 parameters, including \deltamd,
the average mistag fraction and the 
difference between \Bz\ and \Bzb\ mistags for 
each tagging category (8), parameters for the signal 
\deltat\ resolution (16), and parameters for background time dependence (5), 
\deltat resolution (6), and effective dilutions (8).
The value of \deltamd\ was kept hidden throughout the analysis until
all analysis details and
the systematic errors were finalized,
to eliminate possible experimenter's bias.

\begin{table}[!htb]
\begin{center}
\caption{Results for \deltamd\ and a subset of the
parameters obtained from the likelihood fit to the 
\deltat\ distributions. \deltamd\ includes small 
corrections described in the text.}
\label{tab:result-likeli}
\begin{tabular}{lclc}
\hline\hline
Parameter             & Fit Value         &  Parameter        & Fit Value        \\
\hline
\deltamd\             & $0.516 \pm 0.016$ &                   &                  \\
\mistag({\tt Lepton}) & $0.079\pm 0.014$  & \mistag({\tt NT1})& $0.219\pm 0.022$ \\
\mistag({\tt Kaon})   & $0.166\pm 0.012$  & \mistag({\tt NT2})& $0.344\pm 0.020$ \\
$S_1$({\tt Run1})     & $1.37 \pm 0.09$   & $S_1$({\tt Run2}) & $1.19 \pm 0.11$  \\
$f_2$({\tt Run1})     & $0.014\pm 0.010$  & $f_2$({\tt Run2}) & $0.015\pm 0.010$ \\
$f_3$({\tt Run1})     & $0.008\pm 0.004$  & $f_3$({\tt Run2}) & $0.000\pm 0.014$ \\
\hline\hline
\end{tabular}
\end{center}
\end{table}

\begin{figure}[!htb]
\begin{center}
  \includegraphics[width=\linewidth]{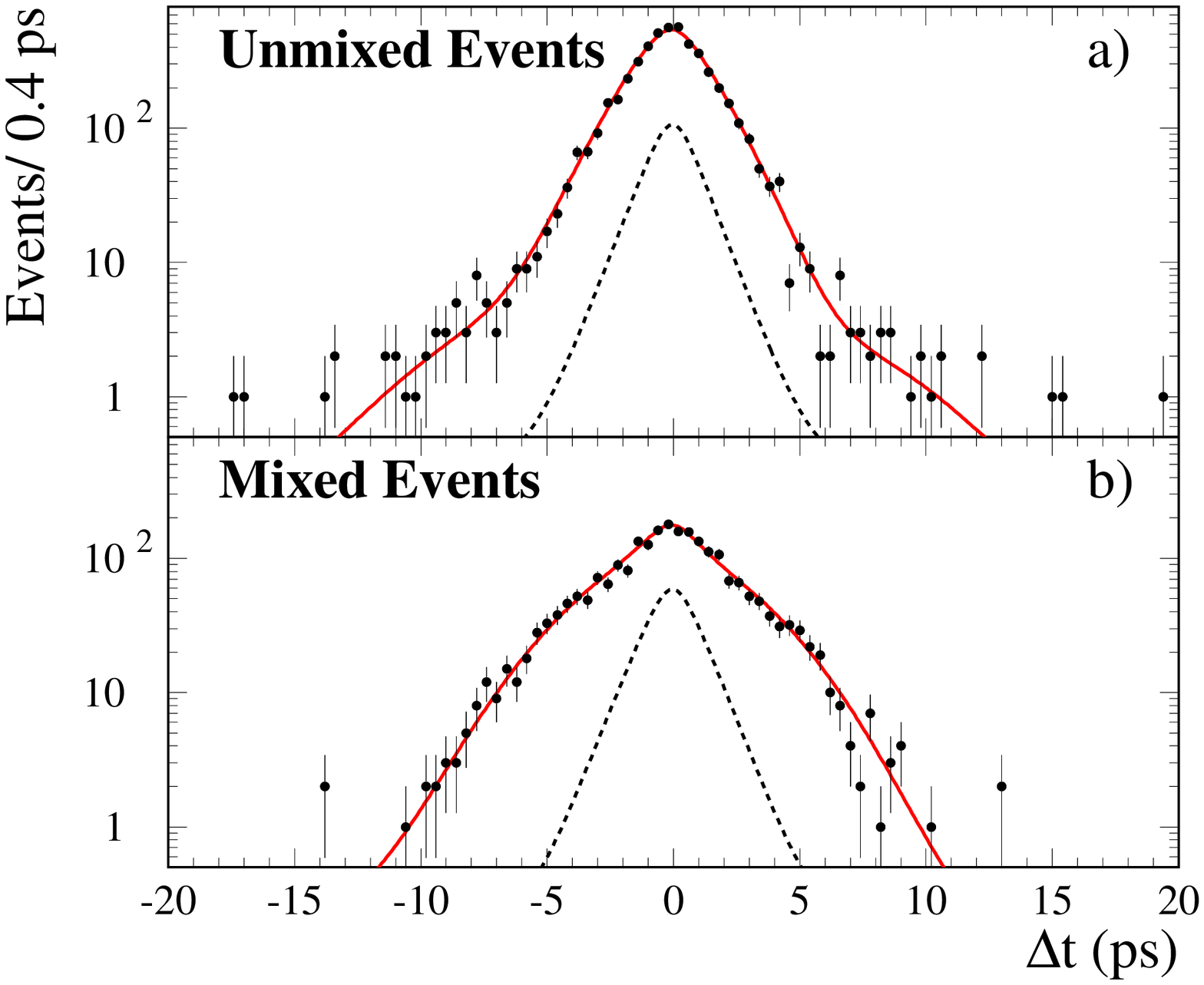}
  \includegraphics[width=\linewidth]{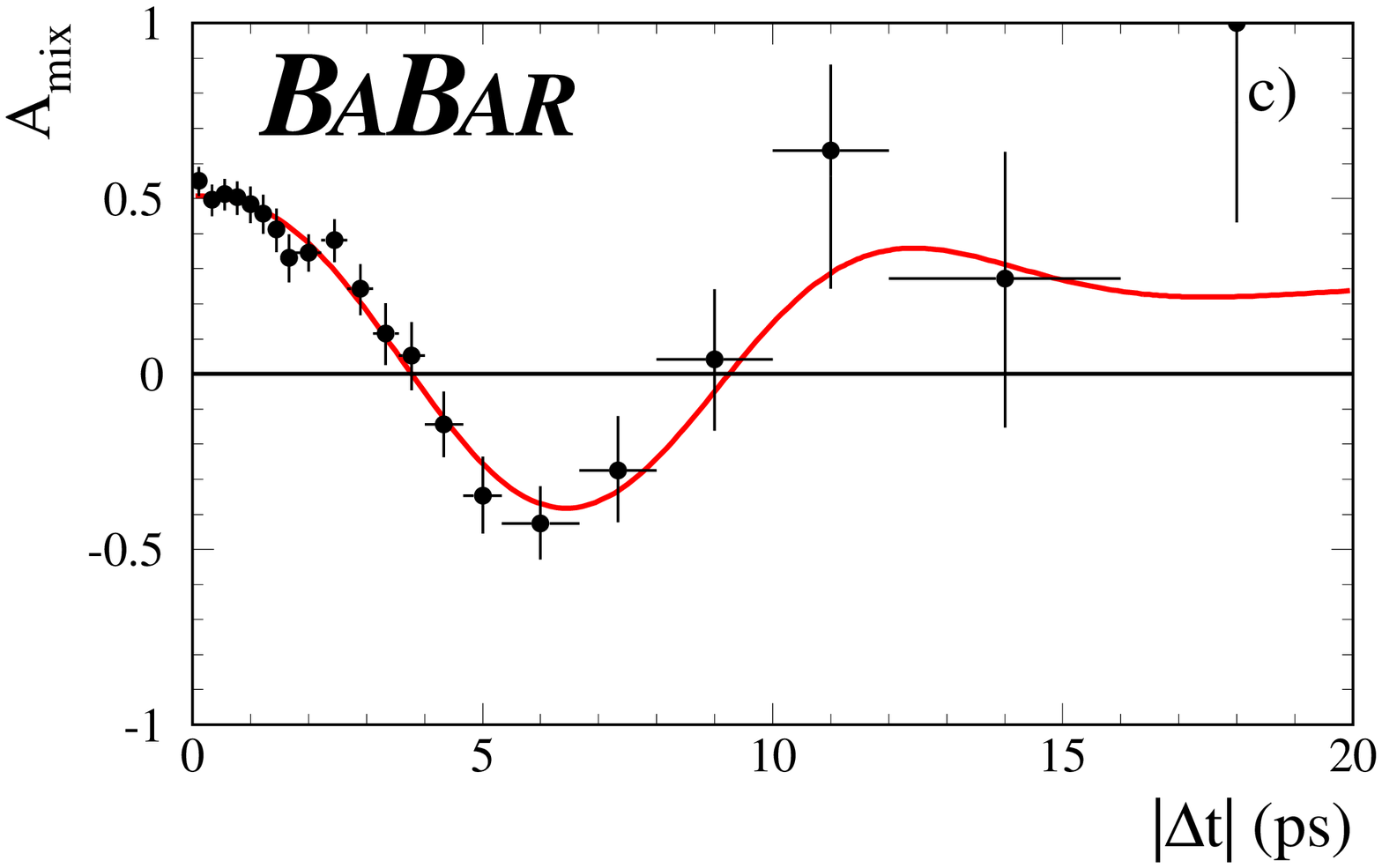}
  \caption{Distributions of \deltat\ in data for the selected a) unmixed 
and b) mixed events ($\mes(B_{\rm rec})>5.27$\gevcc), with projections of the
likelihood fit (solid) and the contribution of the background (dashed) 
overlaid. The time-dependent mixing asymmetry ${\cal A}_{mix}(|\deltat|)$
is shown in c).
    }
\label{fig:deltat}
\end{center}
\end{figure}

The results from the likelihood fit to the tagged \Bz\ sample are
summarized in Table~\ref{tab:result-likeli}.  The probability
to obtain a likelihood smaller than that observed is 44\%, evaluated with a
parameterized Monte Carlo technique.  The value of 
\deltamd\ given by the fit, prior to final corrections, is 
$\Delta m_{d,{\rm fit}} = 0.525 \pm 0.016 \hips$.
One method for displaying the result of the full likelihood fit
is to use the observed mixing asymmetry,
\begin{equation*}
{\cal A}_{mix}(\deltat) = \frac{N_{unmixed}(\deltat)-N_{mixed}(\deltat)}{N_{unmixed}(\deltat)+N_{mixed}(\deltat)}.
\label{eq:asym}
\end{equation*}
If the flavor tagging and \deltat\ determination were perfect,
the asymmetry as a function of 
\deltat\ would be a cosine with unit amplitude.
The amplitude is diluted by the mistag probability and the experimental
resolution for \deltat.
The observed \deltat\ distributions of both the
mixed and unmixed events, and their asymmetry
${\cal A}_{mix}$ are shown along with projections of
the likelihood fit result in Fig.~\ref{fig:deltat}.

Since the parameters of the \deltat\ resolution for both signal and backgrounds
are free parameters in the fit, their contribution to the uncertainty on
\deltamd\ is included as part of the statistical error.
Remaining systematic errors arise from the choice
of the signal \deltat\ resolution description, its capability to
handle outliers and various worst-case SVT misalignment scenarios
($\pm 0.005\hips$), and by
approximations and uncertainties in the $\Delta z$ to $\Delta t$
conversion due to the knowledge of the absolute $z$ scale of the
\babar\ detector and \pep2\ boost (less than $\pm 0.002\hips$).
Systematic errors due to background include the choice of
its \deltat\ distribution and resolution description
($\pm 0.002\hips$), variation of the sum of background fractions
from the separate \mes fits, and
the uncertainty on the magnitude of the small \Bu\ component
of the signal ($\pm 0.002\hips$).
A correction of $-0.002\hips$, derived from data, is made to account 
for the small variation of the background composition as a function of 
\mes, which affects the background \deltat\ distribution.
The statistical error ($\pm 0.002\hips$) on this extrapolation from the 
sideband to the signal region is included as a systematic uncertainty.
An additional correction of $-0.007\hips$  is applied for a bias observed
in fully simulated Monte Carlo events. The bias is mainly due to correlations
between the mistag rate and the \deltat\ resolution that are not explicitly
incorporated into the likelihood function. The systematic error assigned to this
correction includes contributions from the statistical precision of 
the Monte Carlo study
($\pm 0.003\hips$), model variations due to uncertain branching fractions and lifetimes of the
tag-side $D$ mesons and the assumed fraction of wrong-sign kaons produced
in $B$ decays ($\pm 0.001\hips$), and variation of the requirement on the
maximum allowed value of $\sigma_{\deltat}$ ($\pm 0.003\hips$).
Finally, the variation of the fixed 
\Bz\ lifetime within the known errors~\cite{PDG2000} leads to
a systematic uncertainty of $\pm 0.006\hips$.

In conclusion, 
a new technique involving the time-difference distribution of a tagged sample of 
fully-reconstructed neutral \B\ decays has been used to determine
the \Bz-\Bzb\ mixing frequency \deltamd\ to be
\begin{equation*}
\deltamd = 0.516 \pm 0.016 ({stat}) \pm 0.010 ({syst})\,\hips.
\end{equation*}
This is one of the single most precise measurements available.
Moreover, the error on \deltamd\ is still dominated by the size
of the reconstructed \Bz\ sample, leaving substantial room for further
improvement as more data are accumulated. 
The result is consistent with the current world average~\cite{PDG2000}
and a recent \babar\ measurement with a dilepton sample~\cite{BabarPub0119}.
The analysis
shares the same flavor-eigenstate sample as used for the determination
of \stwob, thereby providing an essential
validation for the reported \stwob\ result~\cite{BabarPub0118}.

We are grateful for the excellent luminosity and machine conditions
provided by our \pep2\ colleagues.
The collaborating institutions wish to thank 
SLAC for its support and kind hospitality. 
This work is supported by
DOE
and NSF (USA),
NSERC (Canada),
IHEP (China),
CEA and
CNRS-IN2P3
(France),
BMBF
(Germany),
INFN (Italy),
NFR (Norway),
MIST (Russia), and
PPARC (United Kingdom). 
Individuals have received support from the Swiss NSF, 
A.~P.~Sloan Foundation, 
Research Corporation,
and Alexander von Humboldt Foundation.


\begin{thebibliography}{99}
\bibitem{conjugate}
The symbol \Bz\ refers
to the $B_d$ meson; charge conjugate
modes are implied throughout this paper.
\end{thebibliography}
\end{document}